\documentclass[aps,prc,twocolumn,showpacs,groupedaddress,floatfix]{revtex4}
\usepackage{graphicx}
\usepackage{dcolumn}
\usepackage{bm}
\begin{document} 
\title{Threshold Electrodisintegration of $^3$He}
\author{R. S. Hicks}
\author{A. Hotta}
\author{S. Churchwell}
\altaffiliation[Present address: ]{Duke University, Durham, NC 27708.}
\author{X. Jiang}
\altaffiliation[Present address: ]{Rutgers, State University of New Jersey, Piscataway, NJ 08855.}
\author{G. A. Peterson}
\author{J. Shaw}
\altaffiliation[Present address: ]{GE Global Research Center, Niskayuna, NY 12309.}
\author{B. Asavapibhop}
\author{M. C. Berisso}
\author{P. E. Bosted}
\author{K. Burchesky}
\author{R. A. Miskimen}
\author{S. E. Rock}
\affiliation{Department of Physics, University of Massachusetts, Amherst, Massachusetts, 01003}
\author{I. Nakagawa}
\altaffiliation[Present address: ]{Jefferson Lab., Newport News, VA 23606.}
\author{T. Tamae}
\author{T. Suda}
\altaffiliation[Present address: ]{The Institute of Physical and Chemical Research, Wako 351-0198, Japan}
\affiliation{Laboratory of Nuclear Science, Tohoku University, Sendai, 982-0826, Japan}
\author{J. Golak}
\author{R. Skibi\'nski}
\author{H. Wita\char'40la}
\affiliation{Institute of Physics, Jagellonian University, PL-30059, Krac\'{o}w, Poland}
\author{F. Casagrande}
\author{W. Turchinetz}
\affiliation{MIT-Bates Linear Accelerator Center, Middleton, Massachusetts, 01949}
\author{A. Cichocki}
\author{K. Wang}
\affiliation{Department of Physics, University of Virginia, Charlottesville, Virginia, 22901}
\author{W. Gl\"{o}ckle}
\affiliation{Institut f\"{u}r Theoretische Physik II, Ruhr Universit\"{a}t Bochum, D-44780 Bochum, Germany}
\author{H. Kamada}
\affiliation{Dept. of Physics, Faculty of Engineering, Kyushu Institute of Technology, 1-1 Sensuicho, Tobata, Kitakyushu 804-8550, Japan}
\author{T. Kobayashi}
\affiliation{College of Radiology, Teikyo University, Tokyo, 173-8605, Japan}
\author{A. Nogga}
\affiliation{Department of Physics, University of Arizona, Tucson, Arizona 85721}

\date{\today}

\begin{abstract}
Cross sections were measured for the near-threshold electrodisintegration of $^{3}$He at momentum transfer values of $q=2.4$, $4.4$, and $4.7$ fm$^{-1}$. From these and prior measurements the transverse and longitudinal response functions $R_T$ and $R_L$ were deduced. Comparisons are made against previously published and new non-relativistic $A=3$ calculations using the best available \textit{NN} potentials. In general, for $q<2$ fm$^{-1}$ these calculations accurately predict the threshold electrodisintegration of $^3$He. Agreement at increasing $q$  demands consideration of two-body terms, but discrepancies still appear at the highest momentum transfers probed, perhaps due to the neglect of relativistic dynamics, or to the underestimation of high-momentum wave-function components.
\end{abstract}
\maketitle

\section{Introduction}
During the past decade much experimental and theoretical progress has been made in the study of the trinucleon system---the first non-trivial test of the adequacy of phenomenological $NN$ potentials. Especially instructive tests are provided by photo- and electro-disintegration reactions. For example, Carlson \textit{et al.}\cite{Carlson02} recently reviewed the status of the $^{3}$He quasielastic response functions measured in inclusive electron scattering far from the elastic scattering peak. In this complementary work we report upon the status of the kinematic region near the break-up threshold of $5.5$ MeV. 

In the early seventies the electrodisintegration of the deuteron near break-up threshold was recognized\cite{Hockert73} as a decisive test of the understanding of meson exchange in the traditional picture of the \textit{NN} force. This reaction is unusually informative because the wave functions of the initial and final states are relatively simple and well-known and, if the electron is deflected to far-backward angles, the break-up is dominated by a pure \textit{M}1, $\Delta T=1$ transition. The contribution of meson exchange currents (MEC) generally grows with increasing three-momentum transfer $q$: at $q=2.5$ fm$^{-1}$ MEC raise the threshold cross section by about a factor of $3$; near $q=3.5$ fm$^{-1}$ MEC account for nearly 100\% of the transverse cross section due to destructive interference between the one-body transition amplitudes. 

Although the threshold electrodisintegration of $^{3}$He aroused similar interest, it took another three decades before Viviani \textit{et al.}\cite{Viviani00} were finally able to confirm the importance of MEC in the trinucleon break-up. Earlier, Hadjimichael \textit{et al.}\cite{Hadjimichael83} had established the need for MEC in the elastic cross sections, but due to the requirement of knowing not just the ground state wave functions, but also those of the continuum, the break-up poses a more challenging test. Following quickly on the paper by Viviani \textit{et al.}, additional evidence for MEC was given in two letters\cite{Xu00,Xiong01} reporting asymmetry measurements for longitudinally-polarized electrons scattered from a polarized $^{3}$He target. The first\cite{Xu00} of these measurements was performed near the quasielastic peak, where MEC effects are small. Stronger evidence for MEC was given in a subsequent paper\cite{Xiong01} on the threshold region, where measurements at $q=1.60$ and $2.27$ fm$^{-1}$ were presented. Near threshold the effect of MEC on the spin-dependent asymmetry is calculated to be large, and although the measurements strongly support this prediction, the agreement is not exact. On the other hand, spin-dependent asymmetries represent an especially demanding test of nuclear theory. 

The delayed confirmation of significant MEC effects in the trinucleon break-up stems from the recent parallel developments in precise empirical \textit{NN} potentials and powerful theoretical methods for calculating exactly the $A=3$ wave function. These requirements have now been met, with such success that we can now claim a detailed understanding of most of the basic properties of the trinucleon, at least at low-to-moderate energies and momenta. 

The advances in \textit{NN} potentials were allowed by high-quality measurements and analyses of \textit{pp} and \textit{np} scattering. Precise non-relativistic potentials\cite{Stoks94,Wiringa95,Machleidt96} were constructed that fit the vast databases with $\chi^2$-per-datum values close to unity. In addition to the usual charge-independent parts, charge-dependence and asymmetry terms were introduced to account for \textit{pp} and \textit{np} scattering simultaneously. The electromagnetic parts of these potentials contain Coulomb, Darwin-Foldy, vacuum polarization, and magnetic moment terms with finite-size properties. Although the calculations shown in this paper rely upon just one\cite{Wiringa95} of these potentials, the Argonne AV18, for the properties investigated here little sensitivity would be expected to the differences between these modern potentials. 

The theoretical techniques devised to solve the three-body Schr\"{o}dinger equation are described in two comprehensive reviews\cite{Glockle96,Carlson98}. Monte Carlo methods, Faddeev techniques, and variational procedures that utilize correlated hyperspherical harmonics have all been successfully employed. Because theoretical predictions for the trinucleon electrodisintegration are sensitive\cite{Meijgaard89} to final state interactions, precise representations are needed not only for the ground state, but also for the final continuum states. 

The calculations of Viviani \textit{et al.}\cite{Viviani00}, made within the pair-correlated hyperspherical harmonics scheme, use the AV18 two-nucleon potential, supplemented by the Urbana-IX three-nucleon interaction\cite{Pudliner95}. Because the calculations assume a \textit{pd} final state, they are confined to the narrow excitation region between the two- and three body break-up thresholds at $E_x=5.5$ and $7.7$ MeV. Both longitudinal and transverse response functions were calculated for $^{3}$He at $q= 0.88$, $1.64$, and $2.47$ fm$^{-1}$, three-momentum transfer values that correspond to experimental results obtained by Retzlaff \textit{et al.}\cite{Retzlaff94} at the MIT-Bates accelerator. The effect of MEC is largest at the highest momentum transfer, $q=2.47$ fm$^{-1}$, where the predicted transverse response function $R_T$ is doubled by including MEC. Even though the experimental points have $\approx 35$\% uncertainties, the scale of the MEC enhancement is so large that the importance of exchange currents is firmly established. 

In this study we present: (1) A new measurement of the threshold transverse response function of $^{3}$He, made at the highest momentum transfer probed by Retzlaff \textit{et al.}, but with uncertainties that provide a more rigorous test of the theoretical predictions; (2) Theoretical calculations that include both two- and three-body break-up. These include a self-consistent treatment of final-state interactions and exchange currents; (3) Additional inclusive scattering measurements at $q=4.4$ and $4.7$ fm$^{-1}$, a kinematic region sometimes speculated to mark the onset of the transition to quark-gluon dynamics; and (4) A new appraisal of the threshold longitudinal response function, facilitated by the improved information on the transverse. 

\section{Experiment and Analysis}
The new measurements were made at the MIT-Bates Linear Accelerator Center at electron scattering angle $\theta=160^\circ$, an angle where the cross section is dominantly transverse. Useful continuum data were obtained at effective incident beam energies of $E_0=263$, $506$, and $549$ MeV, corresponding to three-momentum transfers at the two-body break-up threshold of $q=2.4$, $4.4$, and $4.7$ fm$^{-1}$. Many of the details of this experiment have been previously published in a report on the $^{3}$He elastic magnetic form factor\cite{Nakagawa01}. To recapitulate, the target system contained 4000 STP liters of $^{3}$He cooled to $23$ K and pressurized to $50$ atm. In order to mitigate variation in the $^{3}$He density due to beam heating, the gas flow was highly turbulent, an enlarged beam spot was used, and the beam current was held constant at $19\pm 1\ \mu$A. Scattered electrons were detected in a magnetic spectrometer system that included drift chambers for trajectory information, a gas Cerenkov detector and lead-glass shower counter for particle identification, and three layers of plastic scintillators for triggering and timing. 

\begin{figure}
\includegraphics[scale=0.59]{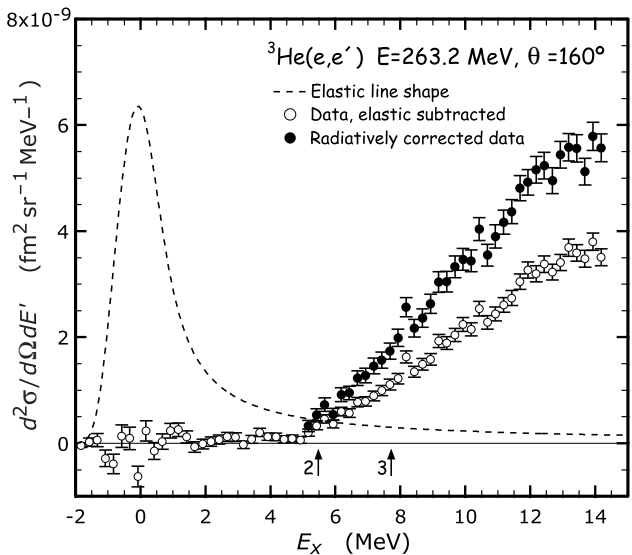}
\caption{\label{fig1} Near-threshold cross section for $^3$He, measured with $263$ MeV incident electrons. The calculated\cite{Nakagawa01} elastic lineshape, indicated by the dashed curve, has been subtracted from the data. The two sets of points show the cross section before and after continuum radiative corrections. Arrows indicate the two- and three-body break-up thresholds. }
\end{figure}

Figure~\ref{fig1} shows the threshold cross section measured with $263$ MeV incident electrons, plotted as a function of excitation energy. Experimental backgrounds have been removed, and corrections applied for dead-time losses and detector inefficiencies. The $^3$He elastic peak and its calculated radiative tail have also been subtracted. As may be seen, in this kinematic region the elastic radiative tail is small. The two- and three-body break-up thresholds are indicated on the figure. Due to the experimental energy resolution, which results mainly from the straggling of electrons travelling different paths within the thick gas target, the continuum cross section begins to rise just before the two-body break-up threshold. 

To expedite the comparison of the data to theoretical predictions, corrections were also made for energy lost by the electron in radiative processes, occurring before, after, or during the primary electro-nuclear interaction. In our experiment the overall effect of these processes is to decrease the cross sections measured near threshold. 

Radiation corrections were applied using the continuum unfolding procedure of Mo and Tsai\cite{Mo69} as implemented by Miller\cite{Miller71}. Even though our target material is low-\textit{Z}, the radiative corrections turn out to be large. In principle, these are precisely calculable, but this requires data more extensive than our limited measurements. Hence we have, in part, had to rely on approximations and models to evaluate the radiative corrections. Nevertheless, the uncertainties in the calculated corrections are expected to be always smaller than the statistical uncertainties of the data. Figure~\ref{fig1} shows the result of radiative unfolding for the spectrum measured at $E_0=263$ MeV; The table lists the radiation-unfolded spectra for all three beam energies. 
\begin{table*}
\caption{\label{tab1} Cross sections and errors (in parentheses) for radiation-unfolded cross sections measured in this work.} 
\begin{ruledtabular}
\begin{tabular}{rl|rl|rl}
\multicolumn{2}{c|}{$E_0=263$ MeV} & \multicolumn{2}{c|}{$E_0=506$ MeV} & \multicolumn{2}{c}{$E_0=549$ MeV}\\
$E_x$ & ${d\sigma}/{d\Omega dE'}$ & $E_x$ & ${d\sigma}/{d\Omega dE'}$ & $E_x$ & ${d\sigma}/{d\Omega dE'}$\\
MeV & pb/sr/MeV & MeV & fb/sr/MeV & MeV & fb/sr/MeV\\
\hline
2.17 & 0.0 (1.4) & 2.58 & 0.6 (1.8) & 2.79 & -0.3 (0.8)\\
2.42 & 0.7 (1.4) & 3.58 & 0.4 (1.3) & 4.29 & -0.2 (0.7)\\
2.67 & 1.9 (1.3) & 4.58 & 0.4 (1.1) & 5.79 & 1.4 (0.9)\\
2.92 & 1.4 (1.3) & 5.58 & 1.5 (1.2) & 7.29 & 3.1 (1.1)\\
3.17 & -0.8 (1.1) & 6.58 & 3.1 (1.4) & 8.79 & 9.1 (1.7)\\
3.42 & 0.9 (1.2) & 7.58 & 5.8 (1.7) & 10.29 & 12.6 (2.0)\\
3.67 & 3.0 (1.2) & 8.58 & 14.5 (2.5) & 11.79 & 16.1 (2.2)\\
3.92 & 2.0 (1.1) & 9.58 & 17.8 (2.8) & 13.29 & 24.0 (2.7)\\
4.17 & 1.5 (1.1) & 10.58 & 27.5 (3.4) & 14.79 & 24.4 (2.8)\\
4.42 & 1.1 (1.0) & 11.58 & 36.5 (3.9) & 16.30 & 36 (3)\\
4.67 & 1.3 (1.0) & 12.58 & 40 (4) & 17.80 & 41 (4)\\
4.92 & 0.8 (1.0) & 13.58 & 54 (5) & 19.30 & 44 (4)\\
5.17 & 3.3 (1.1) & 14.58 & 64 (5) & 20.80 & 47 (4)\\
5.42 & 5.4 (1.2) & 15.58 & 68 (5) & 22.30 & 57 (4)\\
5.67 & 7.3 (1.2) & 16.58 & 84 (6) & 23.80 & 58 (4)\\
5.93 & 5.4 (1.1) & 17.58 & 87 (6) & 25.30 & 60 (5)\\
6.18 & 9.2 (1.3) & 18.59 & 96 (7) & 26.80 & 78 (5)\\
6.43 & 9.5 (1.3) & 19.59 & 101 (7) & 28.30 & 78 (5)\\
6.68 & 12.3 (1.4) & 20.59 & 102 (7) & 29.80 & 92 (6)\\
6.93 & 12.8 (1.4) & 21.59 & 130 (8) & &\\
7.18 & 14.6 (1.5) & 22.59 & 142 (8) & &\\
7.43 & 15.7 (1.5) & 23.59 & 157 (9) & &\\
7.68 & 17.4 (1.6) & 24.59 & 155 (9) & &\\
7.93 & 19.9 (1.6) & 25.59 & 169 (9) & &\\
8.18 & 25.7 (1.8) & 26.59 & 182 (9) & &\\
8.43 & 21.7 (1.7) & 27.59 & 210 (10) & &\\
8.68 & 23.6 (1.7) & 28.59 & 212 (10) & &\\
8.93 & 26.3 (1.8) & 29.59 & 228 (11) & &\\
9.18 & 30.4 (1.9) & & & &\\
9.43 & 30.5 (1.9) & & & &\\
9.68 & 33.3 (2.0) & & & &\\
9.93 & 34.7 (2.0) & & & &\\
10.18 & 34.4 (2.1) & & & &\\
10.43 & 40.4 (2.2) & & & &\\
10.68 & 35.5 (2.1) & & & &\\
10.93 & 39.0 (2.2) & & & &\\
11.18 & 41.7 (2.2) & & & &\\
11.43 & 43.7 (2.3) & & & &\\
11.68 & 48.1 (2.4) & & & &\\
\end{tabular}
\end{ruledtabular}
\end{table*}

\section{Results}
\subsection{Systematics of the transverse response\\ function $R_T$}
In this section we examine the systematic dependence of available experimental information on $R_T$ as a function of incident beam and excitation energies. The radiatively-unfolded, inclusive electron scattering cross sections depend on the longitudinal and transverse response functions according to
\begin{eqnarray*}
\lefteqn{\frac{d^2\sigma\left(E_0,E_x\right)}{d\Omega dE'} = \frac{d^2\sigma_{Mott}}{d\Omega dE'}}\\
 & & \times\left\{R_L\left(q,E_x\right) + \left[\frac{1}{2}+\tan^2\frac{\theta}{2}\right] R_T\left(q,E_x\right)\right\}\ .
\end{eqnarray*}

In the threshold region only Retzlaff \textit{et al.}\cite{Retzlaff94} and K\"{o}bschall \textit{et al.}\cite{Kobschall83} have published separated longitudinal and transverse functions for $^3$He. Nevertheless, due to the large size of the transverse kinematic factor at backward scattering angles, additional information on $R_T$ is provided by the spectra measured at $\theta=180^\circ$ by Jones \textit{et al.}\cite{Jones79} and by our $\theta=160^\circ$ spectra. Chertok \textit{et al.}\cite{Chertok69} published one additional $180^\circ$ spectrum, however this lacks continuum radiative corrections, and, in any case, is superseded by the later measurements of Jones \textit{et al.} made at the same laboratory. Additional spectra were obtained by Kan \textit{et al.}\cite{Kan75}, but at smaller scattering angles where the longitudinal response function is more strongly weighted than the transverse. 

For $180^\circ$ electron scattering the longitudinal response function is negligible and the transverse response can be deduced using
\begin{eqnarray*}
R_T\left(E_0,E_x\right)\approx \left(\frac{2E_0}{\alpha}\right)^2 \cdot \frac{d^2\sigma\left(E_0,E_x\right)}{d\Omega dE'} .
\end{eqnarray*}
Note that the response obtained in this way is given as a function of incident beam energy $E_0$, not the three-momentum transfer $q$. However, for the electron beam energies considered here, $q$ changes slowly in the vicinity of the break-up threshold. 

At the $160^\circ$ angle of our Bates data, longitudinal contributions to the (\textit{e},\textit{e}$'$) cross section are still small. Based on the results of Retzlaff \textit{et al.} and our new calculations, longitudinal contributions to the spectrum measured at $E_0=263$ MeV are $<3.5$\%. These estimated components have been subtracted from the data. No similar allowance has been made for longitudinal contributions to data obtained at $506$ and $549$ MeV, but according to our calculations these are even smaller. 

\begin{figure}
\includegraphics[scale=0.605]{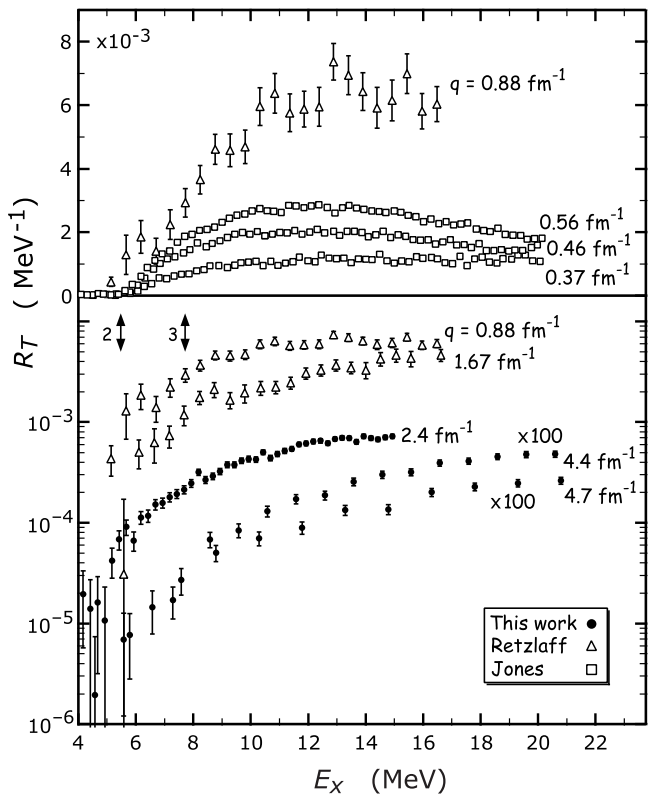}
\caption{\label{fig2} Systematics of near-threshold cross sections measured in three experiments. All spectra are radiatively corrected. The results of Retzlaff \textit{et al.} are for a constant value of three-momentum transfer $q$. As discussed in the text, for the other spectra $q$ changes slowly with $E_x$. For these spectra the indicated $q$-values correspond to the three-body break-up threshold of $7.7$ MeV. }
\end{figure}

Selected results for $R_T$ are compiled as a function of $q$ and $E_x$ in Fig.~\ref{fig2}. As noted above, for the results of Jones \textit{et al.}\cite{Jones79} and our $160^\circ$ Bates experiment, the value of $q$ changes slowly with $E_x$, decreasing by about $0.045$ fm$^{-1}$ between the break-up threshold at $E_x=5.48$ MeV and $E_x=15$ MeV. For these results the $q$-values indicated on the plot were calculated at the three-body break-up threshold of $7.7$ MeV. 

In the threshold region $R_T$ peaks at $q \approx 1$ fm$^{-1}$, and by $q=4.5$ fm$^{-1}$ has decreased by 4 orders of magnitude. For $q<1$ fm$^{-1}$ there is a tendency for $R_T$ to be broadly peaked in the range $E_x=10-20$ MeV. This resembles the distribution of resonant E1 strength seen in photoabsorption measurements, and which is convincingly explained by very recent Faddeev calculations\cite{Golak02} that use the AV18 \textit{NN} potential. Indeed, E1 strength will also be large in inclusive electron scattering at $q<1.5$ fm$^{-1}$, although, at low-$q$ multipoles other than E1 are predicted to make sizeable near-threshold contributions. For example, early calculations of two-body electrodisintegration by Heimbach \textit{et al.}\cite{Heimbach77} indicated considerable M2 strength in the region $E_x<20$ MeV at $q\approx 0.5$ fm$^{-1}$. Additional smaller contributions were obtained from the M1 and M3 multipoles. 

\begin{figure}
\includegraphics[scale=0.68]{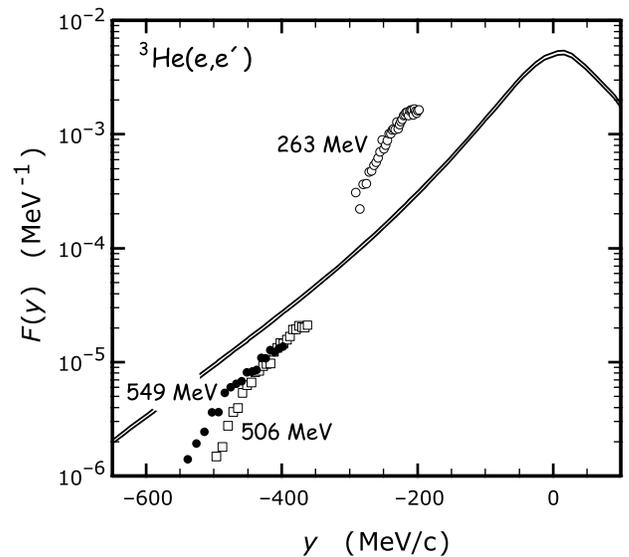}
\caption{\label{fig3} Present data, converted to the quasielastic scaling function $F\left( y\right)$, where $y$ is the initial momentum component parallel to \textbf{q} that would be carried by a quasielastically scattered nucleon\cite{Day90}. Large values of $y$ correspond to large $E_x$. Our results lie outside the indicated narrow band\cite{Day90} corresponding to asymptotic $y$-scaling, indicating that quasifree scattering is not dominant in our kinematic range. From lowest to highest beam energies, the average four-momenta transfer for the data are $Q^2=0.22$, $0.71$, and $0.80$ (GeV/c)$^2$. }
\end{figure}

For $q>2$ fm$^{-1}$, the near-threshold $R_T$ increases monotonically with increasing $E_x$. At still higher momentum transfers quasifree scattering becomes the dominant reaction mechanism and, notwithstanding resonance effects, final-state interactions, and phase-space suppression close to threshold, the monotonic rise seen in the data taken at $E_0=506$ and $549$ MeV has the \textit{appearance} of the high-momentum tail of the quasielastic peak. This tail is of considerable interest since it provides information on elusive high-momentum components of the nuclear wave function\cite{Sick80}. In order to test the quasielastic hypothesis we examined the data to see if the cross section scales with $y$, the initial momentum component parallel to \textbf{q} that would be carried by a quasielastically scattered nucleon. Such scaling is the signature of quasifree scattering\cite{Sick80,Day90}. As shown in Fig.~\ref{fig3}, even though the momentum transfer is sufficiently high, our results lie outside the band that corresponds to asymptotic $y$-scaling. We conclude that, even at the relatively large $q$ of our measurements at $506$ and $549$ MeV, excitation energies of $25$ MeV are insufficient to assure the dominance of quasifree scattering. 

\subsection{Comparison of $R_T$ results with Faddeev calculations that include final-state interactions}
The non-relativistic calculations to which we compare our new spectra are similar to those presented in a previous paper\cite{Golak95}. As described there, bound and continuum \textit{pd} and \textit{ppn} wave functions were obtained by solving Faddeev-like integral equations in momentum space. All final-state interactions are rigorously included. The present calculation is improved in two ways. First, rather than the older-generation Bonn potential, we use the updated Argonne AV18 \textit{NN} interaction, and second, we include MEC contributions, evaluated using Riska's prescription\cite{Riska85}. Most importantly, for these calculations the final-state interactions and exchange currents are fully consistent with the \textit{NN} interaction potential. 

\begin{figure}
\includegraphics[scale=0.60]{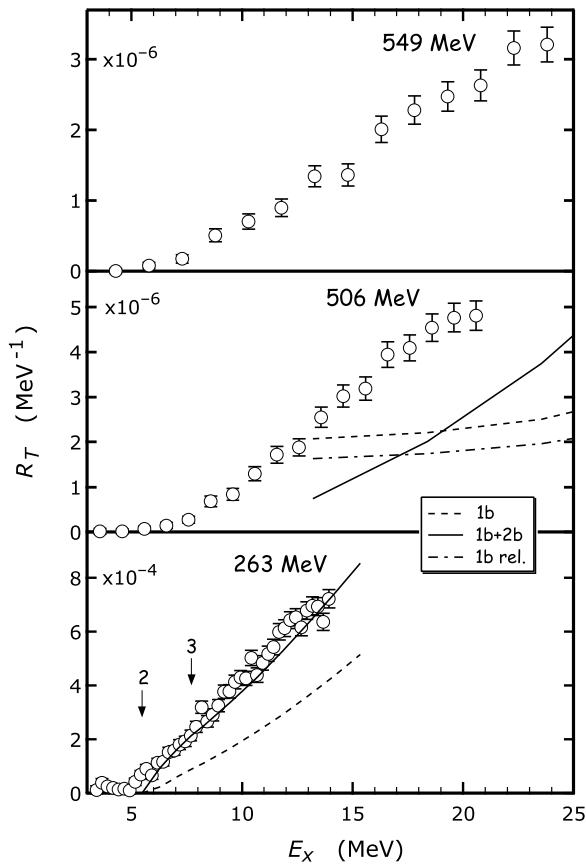}
\caption{\label{fig4} Comparison of present experimental results with our new Faddeev calculations that use the AV18 \textit{NN} interaction. As explained in the text, these calculations include both two- and three-body break-up channels, with a self-consistent treatment of final state interactions. Dashed curve: one-body current only; solid curve: one- and two-body currents; dot-dash: relativistic one-body currents. }
\end{figure}

The theoretical calculations were performed on the Cray SV1 of the NIC in J\"{u}lich, Germany and the NERSC Computational Facility, USA. Despite the computational power of these facilities, the long CPU times required for the calculations limited what could be achieved. The main results are plotted in Fig.~\ref{fig4}. For $263$ MeV the agreement with the data could scarcely be better. These new theoretical predictions may also be compared to the \textit{pd} breakup calculations by Viviani \textit{et al.}\cite{Viviani00} that utilize the same \textit{NN} potential, but which were carried out by means of pair-correlated hyperspherical harmonics, not by solving Faddeev equations. Motivated by the data of Retzlaff \textit{et al.}\cite{Retzlaff94}, Viviani \textit{et al,}'s calculations were performed for a momentum transfer $3.4$\% higher than that of our $263$ MeV measurements. Nevertheless, when this difference is taken into account using the $q$-dependence given by Viviani \textit{et al.}, it is found that the Faddeev and hyperspherical harmonic calculations, both with and without MEC, are almost indistinguishable in the $E_x=5.5-7.7$ MeV range where comparison is valid. It is reassuring to note that our new Faddeev calculation of the three body final state continues to agree well with the data up to the highest excitation energy. According to an earlier calculation\cite{Golak95} based on the previous-generation Bonn B interaction, the three-body contribution to $R_T$ grows relatively slowly above threshold: at $E_x=18$ MeV and $q=0.88$ fm$^{-1}$, it amounts to just half the two-body contribution. 

The agreement of the two calculations, in addition to the agreement with the data, does more than simply confirm the importance of MEC in the $^{3}$He break-up---it underscores how accurate these modern calculations can be---to a point. As indicated in Fig.~\ref{fig4}, the theoretical prediction is less satisfactory for the higher beam energy of $506$ MeV. Here, the relatively flat contribution of the non-relativistic one-body current is modified---in the correct sense---by interference with MEC. Above $E_x=19$ MeV the interference is constructive; below $19$ MeV it is strongly destructive. (This contrasts strongly with the effect of MEC on the $263$ MeV spectrum.) Although this leads to the correct shape, the predicted $R_T$ is about a factor of $2$ too low throughout the threshold region. 

Several factors may account for this discrepancy. For example, the use of non-relativistic dynamics is questionable at such energies. To investigate this we performed an exploratory calculation in which the one-body current only was treated relativistically, and even this in a manner formally inconsistent with the \textit{NN} interaction. Figure~\ref{fig4} shows that this device leads to an even poorer prediction of the data. Moreover, the momentum transfers of our $506$ and $549$ MeV spectra probe small wave function components well beyond the Fermi momentum. These components are usually negligible, but their effects can be magnified in scattering at large $q$, as shown in an analysis of quasifree scattering by Sick, Day, and McCarthy\cite{Sick80}. From this work it was deduced that the ``exact'' $^3$He wave functions obtained from realistic \textit{NN} interactions have high-momentum components that are too small. 

A further indication of theoretical difficulties at large $q$ is evident in the observation\cite{Nakagawa01} that the first diffraction minimum in the elastic magnetic form factor of $^{3}$He is located near $q=4.2$ fm$^{-1}$, somewhat higher than predicted by current theories.
\subsection{$q$-dependence of $R_T$}
Viviani \textit{et al.}\cite{Viviani00} have calculated the $q$-dependence of $R_T$ at a fixed $1.0$ MeV excitation above the \textit{pd} threshold. However, due to the small size of the experimental cross section near threshold, it is unpractical to compare to this prediction. A more reliable comparison may be made by integrating $R_T$ in the range $5.5<E_x<7.7$ MeV, where break-up is confined to the \textit{pd} channel evaluated by Viviani. Our integration of the theory assumes a linear dependence for $R_T$ on $E_x$---an assumption supported by the near-threshold calculations. 

\begin{figure}
\includegraphics[scale=0.68]{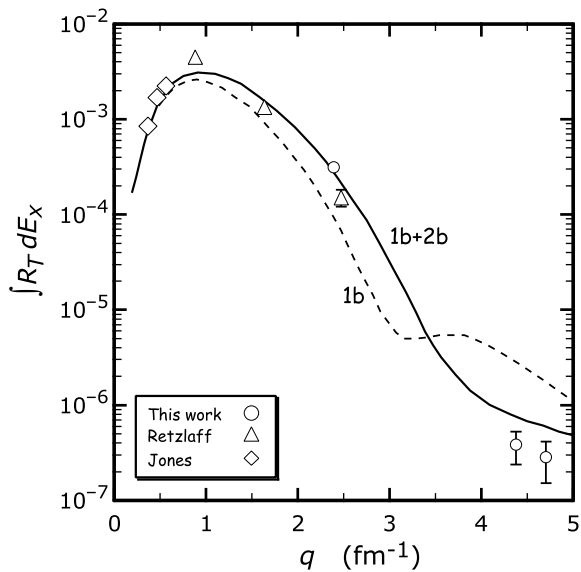}
\caption{\label{fig5} Dependence on three-momentum transfer $q$ of $R_T$, integrated over the range $5.5<E_x<7.7$ MeV. The experimental points are compared with the two-body final state calculations by Viviani \textit{et al.}\cite{Viviani00} Dashed curve: one-body currents only; solid curve: one- and two-body currents. The experimental error bars include systematic uncertainties in the cross sections, as well as uncertainties in $E_x$. Where not shown, the errors are comparable to the size of the points. Additional experimental information on $R_T$ was obtained by K\"{o}bschall \textit{et al.}\cite{Kobschall83}, but close to threshold these results have very large uncertainties. }
\end{figure}

The results are shown in Fig.~\ref{fig5}. With the exception of the two high-$q$ points, excellent quantitative agreement is again obtained. As previously noted, the MEC contribution grows at higher $q$. For $q<3.4$ fm$^{-1}$ MEC terms interfere \textit{constructively} with the one-body matrix elements, raising the prediction by up to a factor of $3$. At higher $q$ the interference is \textit{destructive}, an interpretation supported by the two high-$q$ points, irrespective of the lack of exact quantitative agreement: near $q=4.5$ fm$^{-1}$ the data lie far below the one-body prediction. 

The change in the interference from constructive to destructive agrees with our new theoretical predictions, as indicated in Fig.~\ref{fig4}. But note that the destructive interference is confined to low excitation energies: at large $E_x$ the interference remains constructive, a prediction confirmed by the data.
\subsection{Longitudinal response function} 
\begin{figure}
\includegraphics[scale=0.65]{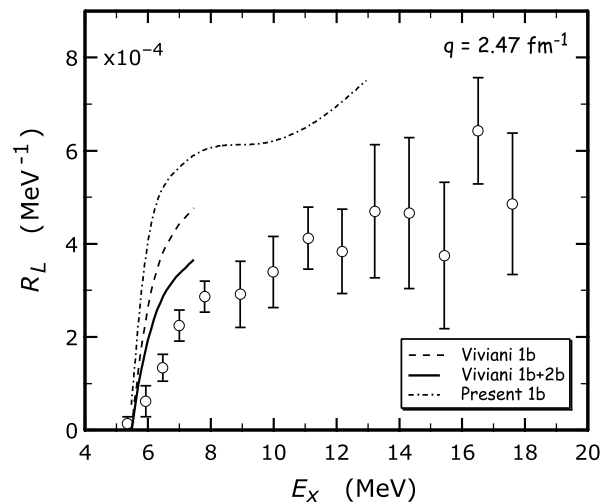}
\caption{\label{fig6} Longitudinal response function of $^3$He, for $q=2.47$ fm-1. The experimental points of Retzlaff \textit{et al.}\cite{Retzlaff94} are compared to the \textit{dp} final state calculations by Viviani \textit{et al.}\cite{Viviani00}. The dashed curve is for the one-body part only, whereas the solid curve includes two-body charge terms. The present calculation is a one-body result for both two- and three-body break-up with full calculation of final state interactions. }
\end{figure}

Unlike the slow and monotonic rise of the transverse response function, the longitudinal response rises abruptly in the first $2$ MeV above threshold\cite{Retzlaff94,Kobschall83,Kan75}, a feature attributed\cite{Kan75} to a $^2S\rightarrow ^2S$ Coulomb monopole transition. This behaviour is illustrated in Fig.~\ref{fig6}, which reproduces the experimental results of Retzlaff \textit{et al.}\cite{Retzlaff94}, obtained at $q=2.47$ fm$^{-1}$. Our calculation for this momentum transfer has the right shape, but exceeds the data by about a factor of $2$. In part, this is attributed to our neglect of the Coulomb barrier that would suppress the emission of low-energy protons. Indeed, as shown in the figure, the calculation of the \textit{pd} electrodisintegration by Viviani \textit{et al.}\cite{Viviani00}, which includes the Coulomb term, lies closer to the data. 

Still better agreement is obtained by including two-body charge operators. As noted by Viviani \textit{et al.}, these operators have relativistic origins and should properly be evaluated by including, in a self-consistent way, relativistic effects in both the interaction models and the nuclear wave functions. Lacking such a method, the only recourse is to perform a model-dependent calculation. This contrasts with the evaluation of the two-body current operators that contribute to the transverse response where, according to the classification scheme of Riska\cite{Riska85}, the main parts of these operators are fixed by current conservation. 

Nevertheless, as Fig.~\ref{fig6} shows, even with the inclusion of the Coulomb and two-body charge terms, the best available theoretical result still exceeds the data by roughly $50$\%. This is a small but notable blemish in what otherwise is a remarkably precise and comprehensive theoretical description of the $^{3}$He threshold photo- and electro-disintegration. It emphasizes the potential value of a more rigorous treatment for the two-body charge operators. Further evidence for this comes from the efforts of Schiavilla and collaborators\cite{Carlson98,Marcucci98}, to predict the charge form factors of $A=3$ and $A=4$ nuclei. 

\begin{figure}
\includegraphics[scale=0.68]{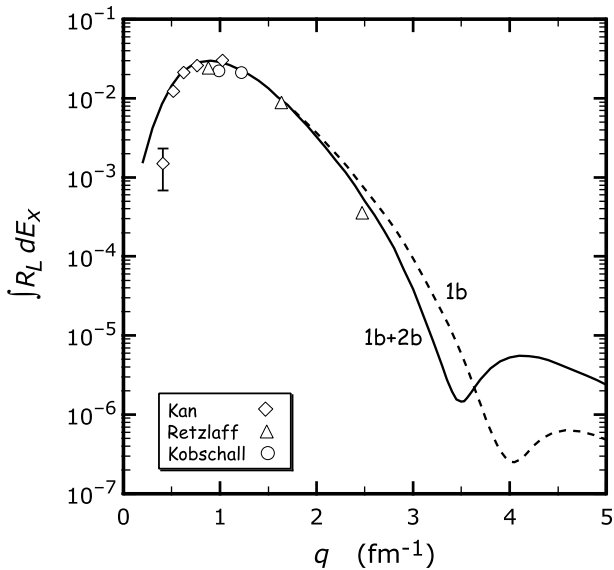}
\caption{\label{fig7} Dependence on three-momentum transfer $q$ of $R_L$, integrated over the range $5.5<E_x<7.7$ MeV. The experimental points are compared with the two-body final state calculations by Viviani \textit{et al.}\cite{Viviani00} Dashed curve: one-body terms only; solid curve: sum of one- and two-body terms. Where not shown, the errors are comparable to the size of the points. }
\end{figure}

Figure~\ref{fig7} shows the threshold $q$-dependence of $R_L$. As for Fig.~\ref{fig5}, we have integrated the experimental and theoretical response functions in the range $5.5<E_x<7.7$ MeV, where break-up is restricted to the \textit{pd} channel. Our integration in this case takes note of the curvature in the dependence of $R_L$ on $E_x$. According to the calculations of Viviani \textit{et al.}\cite{Viviani00}, this diminishes at large $q$. The experimental points in Fig.~\ref{fig7} include the results of Retzlaff \textit{et al.}\cite{Retzlaff94}, K\"{o}bschall \textit{et al.}\cite{Kobschall83}, and Kan \textit{et al.}\cite{Kan75}. Kan \textit{et al.} were unable to extract $R_L$ from their limited measurements, however, by virtue of the now-precise knowledge of $R_T$ at the $q$-values of their measurements, this separation becomes possible. 

As the plot shows, for $q<2$ fm$^{-1}$ the agreement between experiment and theory is generally excellent, while the significance of the discrepancy with Retzlaff's point at $q=2.47$ fm$^{-1}$ has been discussed in some length. The disagreement for the lowest-$q$ point of Kan \textit{et al.} can perhaps be attributed to uncertainties inherent in a $>50$\% transverse subtraction. For other points, the transverse contribution is typically $<20$\%.
\section{Summary}
For $q<3$ fm$^{-1}$, exact non-relativistic calculations using the best available \textit{NN} potentials give a very good description of measurements of the $^{3}$He threshold electrodisintegration. At low momentum transfer one-body matrix elements predominate, but as $q$ increases the two-body contribution grows significant, particularly in the transverse part of the cross section. For example, at $q=2.4$ fm$^{-1}$, corresponding to one of the three new measurements reported here, the inclusion of two-body terms raises the predicted $R_T$ by a factor of two, bringing the theory into close agreement with the data---proof that our understanding of MEC is accurate. 

According to the near-threshold theoretical predictions for $R_T$, at $q\approx 3.5$ fm$^{-1}$ the interference between one- and two-body terms switches from constructive to destructive. This is supported by our two other measurements, made at $q\approx 4.5$ fm$^{-1}$. These points fall about a factor of $7$ below the one-body prediction, but the destructive interference with two-body current terms lowers the prediction. That it still exceeds the data by a factor of two suggests the need for a more complete interference, but other factors cannot be overlooked. 

For example, the high-$q$ measurements are in a kinematic region sensitive not only to relativistic effects, but also to high-momentum wave function components, which, as indicated by quasifree scattering results, may be too small in ``exact'' wave functions obtained from realistic \textit{NN} interactions. A pointed indication of the importance of relativistic effects is found in the analysis of Viviani \textit{et al.}\cite{Viviani00} of the longitudinal response function at $q=2.47$ fm$^{-1}$. Close to threshold there exists a factor-of-two disagreement between the one-body predictions and experimental values of $R_L$. This disagreement is reduced, but not entirely resolved, by considering two-body charge matrix elements. These are equivalent to relativistic corrections. At this time estimates of these corrections are model-dependent: the discovery of a rigorous, self-consistent procedure for evaluating relativistic effects poses a considerable challenge. As has been repeatedly noted, a relativistic formulation of effective hadronic theory is essential for a satisfactory understanding of the transition from hadron to quark regimes. 

Additional measurements are needed to guide the theoretical development, particularly above $q=2$ fm$^{-1}$, where few data currently exist. An upcoming experiment in Hall A at Jefferson Laboratory\cite{Aniol00} is aimed at measuring the elastic form factors of $^{3}$He and $^{4}$He, starting at about $4.5$ fm$^{-1}$. Unfortunately, broad energy resolution will limit what can be learned about the threshold break-up. 

Especially valuable would be a new measurement of the $^{3}$H isobar, for which existing data are very sparse. One useful simplification offered by $^{3}$H is the absence of a Coulomb interaction---difficult to incorporate into Faddeev calculations---between the break-up products.

\begin{acknowledgements}
This work was supported by U.S. DOE Grant DE-FG02-88ER40415, NSF Grants PHY9114958 and PHY0070858, Grant in Aid for International Research (Nos. 04640281 and 08044050) by the Ministry of Education, Science, and Culture in Japan, and by the Polish Committee for Scientific Research (grant No. 2P03B02818). The theoretical calculations were performed on the Cray SV1 of the NIC in J\"{u}lich, Germany and the NERSC Computational Facility, USA.
\end{acknowledgements}


\end{document}